\def\b{\begin{equation}}
\def\e{\end{equation}}
\def\balll{\begin{array}{lll}}
\def\ea{\end{array}}
\def\bea{\begin{eqnarray}}
\def\eea{\end{eqnarray}}

\documentclass[floatfix,amsmath,amssymb]{revtex4}
\usepackage{epsfig}

\usepackage[latin1]{inputenc}
\begin{document}
\title{Quantum radiation force on a moving mirror 
with Dirichlet and Neumann boundary conditions at vacuum, finite temperature and coherent states.}

\author{Danilo T. Alves$^{1}$, Edney R. Granhen$^{1,2}$ and Mateus G. Lima$^{1}$}
\affiliation{(1) - Faculdade de F\'\i sica, Universidade Federal do
Par\'a, 66075-110, Bel\'em, PA,  Brazil\\(2) - Centro Brasileiro de Pesquisas Físicas, Rua Dr. Xavier Sigaud,
150, 22290-180, Rio de Janeiro, RJ, Brazil}

\date{\today}
\begin{abstract}
We consider a real massless scalar field in a two-dimensional spacetime,
satisfying Dirichlet or Neumann boundary condition at the instantaneous position of a moving boundary.
For a relativistic law of motion, we show that
Dirichlet and Neumann boundary conditions yield the same radiation force on a moving mirror when
the initial field state is invariant under time translations. 
We obtain the exact formulas for the energy density of the field and the radiation force on the boundary for 
vacuum, thermal and coherent state. In the nonrelativistic limit, our results coincide with those found
in the literature.
\end{abstract}

\maketitle

%

In the 70s decade the first works investigating
the quantum problem of the radiation emitted by moving mirrors in vacuum
were published, motivated by the investigation of particle creation
in the nonstationary universe (see Refs. \cite{Moore-1970, Dewitt-PhysRep-1975, Fulling-Davies-PRS-1976-I, 
Fulling-Davies-PRS-1977-I, Fulling-Davies-PRS-1977-II, Candelas-Raine-JMP-1976, 
Candelas-Raine-PRS-1977}). 
Fullling and Davies \cite{Fulling-Davies-PRS-1976-I} studied the moving mirror radiation problem 
in the context of a real scalar field in a two dimensional spacetime. They obtained an exact formula for the finite physical part
of the expected value of the energy-momentum tensor, assuming the
initial state as the vacuum. Their results revealed that 
the radiation is originated at the mirror and propagates
away from it. 
Ford and Vilenkin \cite{Ford-Vilenkin-PRD-1982} developed a perturbative method which can be applied to mirrors
moving in small displacements and with nonrelativistic velocities. 
In this approximation, they obtained, for a real scalar field in a two dimensional spacetime, 
that the radiation force is proportional to the third time derivative of the mirror law of motion, which is 
the nonrelativistic limit of the result obtained in Ref. \cite{Fulling-Davies-PRS-1976-I}. Ford and Vilenkin also
applied their method to a scalar field in four-dimensional spacetime, obtaining a force proportional to the fifth time derivative of
the displacement of the mirror.
Davies and Fulling \cite{Fulling-Davies-PRS-1977-II} (see also Ref. \cite{Birrel-Davies-Book-1982}) found a class of hyperbolic
trajectories of a mirror moving in a 1+1 Minkowski spacetime, for which the emitted radiation corresponds to a thermal spectrum,
that could be registered by an inertial detector. 
Moreover, several works have focused on the problem of moving mirrors placed in a thermal bath
(considered as the ``in'' state, instead
of the vacuum). Jaekel and Reynaud \cite{Jaekel-Reynaud-JPI-and-PLA-1993} obtained for the scalar field in $1+1$ 
dimensions the thermal contribution to the dissipative force 
proportional to the velocity of the mirror, valid in the nonrelativistic limit. Thermal effects 
have also been considered in Refs. 
\cite{plunien-PRL-2000, Jaekel-Reynaud-PLA-1993, papers-temperatura, papers-temperatura-II, papers-temperatura-III,Lambrecht-JOptB-2005}. The coherent state is
another initial field state which has been considered \cite{Alves-Farina-Maia-Neto-JPA-2003, estados-coerentes}, 
as well as the superposition of coherent sates used to study decoherence via the dynamical 
Casimir effect \cite{estados-coerentes-superpostos}.
Furthermore, attention has been given to the role that boundary conditions
play on the dynamical Casimir effect. In the static Casimir effect, different
boundary conditions can change the sign of the Casimir force \cite{forca-repulsiva-casimir}.
Applications of this change, for instance, in the building of nanoelectromechanical systems have
been discussed \cite{forca-repulsiva-casimir-II}.
The role of Dirichlet and Neumann conditions on the
static Casimir force has been investigated in Refs. \cite{papel-dirichlet-neumann-forca-estatica-casimir}
in the context of the scalar field.
Recently, several works have investigated the influence of different boundary conditions on 
the dynamical Casimir effect \cite{cas-din-papel-conds-fronteira,Alves-Farina-Maia-Neto-JPA-2003}.
In this context, Farina, Maia Neto and one of the present authors,
showed that Dirichlet and Neumann boundary conditions yield the same force on a moving mirror
when the initial field state is symmetrical under time translations \cite{Alves-Farina-Maia-Neto-JPA-2003}. 
However, the validity of this conclusion, obtained in the context of the Ford-Vilenkin approach \cite{Ford-Vilenkin-PRD-1982}, 
was restricted to the nonrelativistic and small displacement approximations. 

The exact method of Fulling-Davies \cite{Fulling-Davies-PRS-1976-I} is frequently considered
as less flexible than the perturbative methods because it is based on
conformal transformations and applicable only 
for the two-dimensional spacetime \cite{papers-temperatura-II,
papers-temperatura-III}, whereas the perturbative approach can be extended
to higher dimensions. On the other hand, the Fulling-Davies method enables to 
obtain exact results for the scalar field in the two-dimensional spacetime. 
In the present work, instead of following the approximate approach 
considered in Ref. \cite{Alves-Farina-Maia-Neto-JPA-2003}, 
we use the exact approach proposed by Fulling and Davies \cite{Fulling-Davies-PRS-1976-I}, and 
show that, also for relativistic laws of motion, 
Dirichlet and Neumann boundary conditions yield the same radiation force on a moving mirror when
the initial field state is symmetrical under time translations. 
In this context, we calculate for a real massless scalar field in a two-dimensional spacetime the 
exact (relativistic) formulas for the energy density of the field and for the force on the mirror, for Dirichlet and Neumann conditions,
at vacuum, finite temperature and coherent initial states of the field.
In the nonrelativistic limit, these exact formulas lead to the approximate results found in Refs. 
\cite{Jaekel-Reynaud-JPI-and-PLA-1993,Alves-Farina-Maia-Neto-JPA-2003}. 
%

\begin{figure}
\epsfig{file=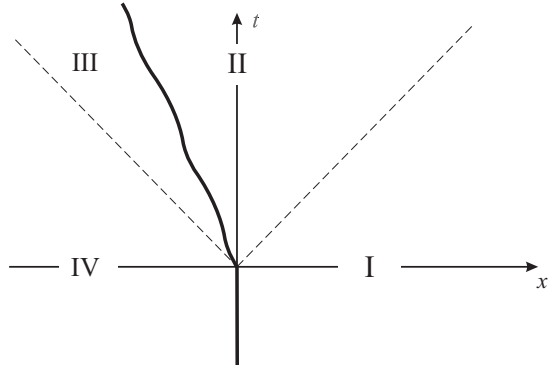,angle=00,width=0.4 \linewidth,clip=}
\caption{Moving mirror trajectory. The dashed lines are null-lines separating the regions
I from II, and III from IV.}
\label{trajetoria-fronteira}
\end{figure}

We start by reviewing the exact field solution 
for Dirichlet and Neumann dynamical boundary conditions
\cite{Dewitt-PhysRep-1975, Fulling-Davies-PRS-1976-I, Fulling-Davies-PRS-1977-I}. 
Let us consider the field satisfying the Klein-Gordon equation
(we assume throughout this paper $\hbar=c=1$):
$
\left(\partial _{t}^{2}-\partial _{x}^{2}\right) \phi \left(
t,x\right) =0,
$
and obeying Dirichlet ($\phi ^{\prime }(t^{\prime},x^{\prime})\vert_{boundary}=0
$) or Neumann ($\partial _{x^{\prime }}\phi ^{\prime }(t^{\prime},x^{\prime})\vert_{boundary}=0$
) condition, taken in the instantaneously co-moving Lorentz frame,
at the moving boundary position $x=z(t)$. We examine 
the particular set of mirror trajectories for which $z(t<0)=0$, as shown
in Fig. \ref{trajetoria-fronteira}. 
Using the appropriate Lorentz transformation, Dirichlet
and Neumann conditions can be written in terms of quantities in the laboratory inertial frame
as follows:
{
\small
$
\phi \left[ t,z(t)\right]=0
$} 
and
{
\small
$
\left. \left\{ \left[ \dot{z}(t){\partial_t}+{\partial_x}\right] \phi \left( t,x\right) \right\} \right|_{x=z(t)}=0,
$} 
respectively. The field solution
can be obtained by exploiting the conformal invariance of the Klein-Gordon equation. Considering
the conformal coordinate transformations
$
t-x=f\left( w-s\right)
$
and
$
t+x=g\left( w+s\right),
$
the scalar wave equation in 1+1 dimensions remains invariant:
$
\left({\partial_{w}^{2}}-{\partial_{s}^{2}}\right) \phi \left( w,s\right) =0.  \label{Klein-Gordon-w-s}
$
Functions $f$ and $g$ can be set up, for a large class of 
laws of motion $z(t)$, so that the curve $s=0$ coincides 
with the boundary trajectory $x=z(t)$: $[t,z(t)]\rightarrow(w,0)$ \cite{Fulling-Davies-PRS-1976-I}.
Then:
{
\small
$
\phi \left[t,z(t)\right]=0\rightarrow \phi(w,0)=0
$} 
and
{
\small
$
\left. \left\{ \left[ \dot{z}(t){\partial_t}+{\partial_x}\right] \phi \left( t,x\right) \right\} \right|_{x=z(t)}=0
\rightarrow [{\partial_s }\phi(w,s)]_{s=0}=0.
$} 
The mode solution of the wave equation with static Dirichlet or Neumann boundary conditions in $(w,s)$ space 
are well known so that, coming back to $(t,x)$ coordinates, we get:
$
\hat{\phi}\left( t,x\right) =\int_{0}^{\infty }d\omega \left[ \hat{a}%
_{\omega }\phi _{\omega }+\hat{a}_{\omega }^{\dag }\phi ^{\ast }\right],
$
where
\begin{equation}
\phi _{\omega }\left( t,x\right) =\left( 4\pi \omega \right) ^{-\frac{1}{2}}%
\left[ \gamma e^{-i\omega \pi r\left(v\right) }+ \gamma^{*}e^{-i\omega \pi
p\left( u\right) }\right]
\label{field-solution}
\end{equation}
form a complete set of positive-frequency solutions, and
$u=t-x$, $v=t+x$. In Eq. (\ref{field-solution}), we introduce a notation
which enables us to put into a single formula the solutions for Dirichlet and Neumann
boundary conditions, and also the solutions for the right hand side (regions I and II in Fig.\ref{trajetoria-fronteira})
and the left side (regions III and IV) of the mirror. In this sense, 
for $\gamma=1$, Eq. (\ref{field-solution}) 
gives the Neumann solution, whereas for $\gamma=i$ we have
the solution for Dirichlet boundary condition. 
For the regions I and II showed in Fig. \ref{trajetoria-fronteira}:
$r\left(v\right)=v$ and $2\tau(u)-u=f^{-1}\left( u\right) \equiv p\left( u\right)$,
where $\tau(u)$ can be obtained from $\tau(u)-z[\tau(u)]=u$; for the regions
III and IV: $p\left(u\right)=u$ and 
$2\tau(v)-v=g^{-1}\left( u\right) \equiv r\left( v\right)$, where
$\tau(v)+z[\tau(v)]=v$. As causality requires, the field in the regions I and IV,
represented by $\phi_{0}$, is not affected by the boundary motion \cite{Fulling-Davies-PRS-1976-I},
so that $p$ and $r$ are also chosen to be identity functions in these static regions. 
Hereafter we consider the averages $\langle...\rangle$ taken over an arbitrary initial field state
(regions I and IV) assumed here, for simplicity, as being
the same one for both sides of the mirror. 
We continue our analysis writing the correlator function ${\cal C}=\left\langle \phi _{0}(t,x)\phi _{0}(t^{\prime },x^{\prime
})\right\rangle$ split in the following manner:
${\cal C}={\cal C}_{vac}+
{\cal C}_{\left\langle \hat{a}^{\dag}\hat{a}\right\rangle}+
{\cal C}_{\left\langle \hat{a}\hat{a}\right\rangle}$, 
where
$$
{\cal C}_{vac}=\int_{0}^{\infty }d\omega
{\cal F}_1(\omega,\omega^{\prime},|\gamma|,\gamma)|_{\omega=\omega^{\prime}}, 
$$
$$
{\cal C}_{\left\langle \hat{a}^{\dag}\hat{a}\right\rangle}=\int_{0}^{\infty }\int_{0}^{\infty }
{d\omega d\omega ^{\prime }}\left\langle \hat{a}_{\omega
^{\prime }}^{\dag }\hat{a}_{\omega }\right\rangle {\cal F}_1(\omega,\omega^{\prime},|\gamma|,\gamma)+\mbox{c.c.},
$$
$$
{\cal C}_{\left\langle \hat{a}\hat{a}\right\rangle}=\int_{0}^{\infty }\int_{0}^{\infty }{d\omega d\omega ^{\prime }}\left\langle \hat{a}_{\omega
^{\prime}}\hat{a}_{\omega}\right\rangle{\cal F}_1(\omega,-\omega^{\prime},\gamma,|\gamma|)+\mbox{c.c.}
$$
and
\begin{eqnarray}
{\cal F}_1(\omega,\omega^{\prime},\rho,\lambda)&=&e^{-i\left( \omega
t-\omega ^{\prime }t^{\prime }\right) }/(4\pi \sqrt{\omega \omega ^{\prime }})\left[\rho^{2}e^{-i\left(
\omega x-\omega ^{\prime }x^{\prime }\right) }\right.\nonumber
\\&&
\left.+\lambda^{2}e^{-i\left( \omega
x+\omega ^{\prime }x^{\prime }\right) }+\mbox{c.c.}\right]\nonumber.
\end{eqnarray}
From these equations, we see that, in the presence of the boundaries,
${\cal C}_{vac}, {\cal C}_{\left\langle \hat{a}^{\dag}\hat{a}\right\rangle}$ and $ 
{\cal C}_{\left\langle \hat{a}\hat{a}\right\rangle}$ are not symmetric 
under space translations. 
The term ${\cal C}_{vac}$ is symmetric under time translation, 
whereas ${\cal C}_{\left\langle \hat{a}^{\dag}\hat{a}\right\rangle}$
is symmetric if $\left\langle \hat{a}_{\omega
^{\prime }}^{\dag }\hat{a}_{\omega }\right\rangle\propto\delta(\omega
^{\prime }-\omega)$. On the other hand ${\cal C}_{\left\langle \hat{a}\hat{a}\right\rangle}$,
if non-null, is not symmetric under time translations. In the context of the perturbative
approach \cite{Ford-Vilenkin-PRD-1982}, the radiation force can be given in terms of correlation functions depending on the unperturbed
field operador $\phi_0$: $\left[\partial_x\partial_{x^{\prime}}{\cal C}\right]_{x^{\prime}=x=0}$
and $\left[{\cal C}\right]_{x^{\prime}=x=0}$ for Dirichlet and Neumann
boundary conditions respectively \cite{Ford-Vilenkin-PRD-1982, Alves-Farina-Maia-Neto-JPA-2003}. 
As shown in Ref. \cite{Alves-Farina-Maia-Neto-JPA-2003}, the parts of the force related to 
${\cal C}_{vac}$ and ${\cal C}_{\left\langle \hat{a}^{\dag}\hat{a}\right\rangle}$ are the
same for Dirichlet and Neumann boundary conditions. On the other hand, the part related
to ${\cal C}_{\left\langle \hat{a}\hat{a}\right\rangle}$ is different by a sign. 
In other words, in Ref. \cite{Alves-Farina-Maia-Neto-JPA-2003} it is shown that the
difference between the force acting on Dirichlet and Neumann
emerges, in the context of the non-relativistic mirror motion
with small amplitude, from the part of the non-perturbed 
correlator $\cal C$ which is non-invariant under time translations.

Now we will investigate this problem in the context of an exact approach,
starting from the field solution (\ref{field-solution}) and calculating the exact formulas for the expected value of the energy 
density operator
${\cal T}=\langle\hat{T}_{00}(t,x)\rangle$, and the net force $F\left( t\right)$ 
acting on the moving boundary defined by (since $T_{00}=T_{11}$ in this model) :
$F\left( t\right)={\cal T}\left[
t,z(t)\right]^{\left( -\right)
}-{\cal T}\left[ t,z(t)\right]^{\left( +\right) }$,
where the superscript ``+" indicates the regions
I and II, whereas ``-" indicates the regions III and IV in Fig. \ref{trajetoria-fronteira}. 
Let us split ${\cal T}$, writing ${\cal T}={\cal T}_{vac}+
{\cal T}_{\left\langle \hat{a}^{\dag}\hat{a}\right\rangle}+
{\cal T}_{\left\langle \hat{a}\hat{a}\right\rangle}$, 
where:
$$
{\cal T}_{vac}=
1/2\int_{0}^{\infty
}d\omega {\cal F}_2(\omega,\omega^{\prime},|\gamma|,|\gamma|)|_{\omega=\omega^{\prime}},
$$
$$
{\cal T}_{\left\langle \hat{a}^{\dag}\hat{a}\right\rangle}=\int_{0}^{\infty }\int_{0}^{\infty }
d\omega d\omega^{\prime}\langle \hat{a}_{\omega ^{\prime}}^{\dag}\hat{a}_{\omega }\rangle 
{\cal F}_2(\omega,\omega^{\prime},|\gamma|,|\gamma|),
$$
$$
{\cal T}_{\left\langle \hat{a}\hat{a}\right\rangle}=-1/2\int_{0}^{\infty }\int_{0}^{\infty }
d\omega d\omega^{\prime}\langle \hat{a}_{\omega ^{\prime}}\hat{a}_{\omega }\rangle 
{\cal F}_2(\omega,\omega^{\prime},\gamma,\gamma^{*})+{\mbox c.c.},
$$
and
\begin{eqnarray}
{\cal F}_2(\omega,\omega^{\prime},\rho,\lambda)&=&\frac{\sqrt{\omega\omega^{\prime}}}{2\pi}\left\{
\rho^2\left[ r^{\prime }\left( v\right) \right] ^{2}e^{-i(\omega-\omega^{\prime}) r\left(
v\right) }\right. \nonumber
\\
&&+\left. \lambda^2\left[ p^{\prime }\left(u\right) \right] ^{2}e^{-i(\omega-\omega^{\prime}) p\left(
u\right) }\nonumber
\right\}.
\end{eqnarray}
The first term ${\cal T}_{vac}$  is the local energy density related to the vacuum state, which is divergent. 
After regularization and renormalization (see Ref. \cite{Fulling-Davies-PRS-1976-I}),
${\cal T}_{vac}$ can be redefined as the renormalized local energy density:
\begin{eqnarray}
{\cal T}_{vac}&=&-{|\gamma|^2}/{24\pi }\left[ {p^{\prime \prime
\prime }\left( u\right) }/{p^{\prime }\left( u\right) }-({3}/{2}){
p^{\prime \prime }\left( u\right)^{2}}/{p^{\prime }\left( u\right)^{2}}\right]
\nonumber\\
&&+p(u)\rightarrow r(v),  \label{tensor-valor-esperado-vacuo}\nonumber
\end{eqnarray}
where the object appearing in the brackets is known as the Schwarzian derivative.
We remark that, in our notation, for the right side of the mirror we have $r(v)=v$, so that
this equation gives back the formula
found in Ref. \cite{Fulling-Davies-PRS-1976-I} for Dirichlet, and in addition 
we verify that the same formula is valid for the Neumann boundary condition.
Now we investigate the net radiation force $F$ acting on the moving mirror. Let us consider
$F=F _{vac}+F_{\left\langle \hat{a}^{\dag}\hat{a}\right\rangle}+F_{\left\langle \hat{a}\hat{a}\right\rangle}$,
where:
\begin{eqnarray}
F _{vac}&=&|\gamma|^2\left( 1+\dot{z}^{2}\right)
\nonumber\\
&&\times
\left[({\ddot{z}^{2}\dot{z}}/{2\pi })
/{\left( 1-\dot{z}^{2}\right) ^{4}}+({
\dddot{z}}/{6\pi })/{\left( 1-\dot{z}
^{2}\right) ^{3}}\right]\nonumber,
\end{eqnarray}
$$
F_{\left\langle \hat{a}^{\dag}\hat{a}\right\rangle}
=2\int_{0}^{\infty }\int_{0}^{\infty }d\omega d\omega^{\prime }\left	\langle \hat{a}_{\omega ^{\prime
}}^{\dag }\hat{a}_{\omega }\right\rangle {\cal F}_3(\omega,\omega^{\prime},|\gamma|,|\gamma|)
+\mbox{c.c.}\;,  \nonumber
$$
$$
F_{\left\langle \hat{a}\hat{a}\right\rangle}
=-\int_{0}^{\infty }\int_{0}^{\infty }d\omega d\omega^{\prime }\left\langle \hat{a}_{\omega }\hat{a}_{\omega^{\prime
} }\right\rangle {\cal F}_3(\omega,\omega^{\prime},\gamma^{*},\gamma)
+\mbox{c.c.}\;,  \nonumber
$$
and
\begin{eqnarray}
{\cal F}_3(\omega,\omega^{\prime},\rho,\lambda)
&=&\frac{\sqrt{
\omega \omega ^{\prime }}}{4\pi }\left\{-\rho^2\dfrac{
\left( 1+\dot{z}\right)^{2} }{\left( 1-\dot{z}\right)^{2}}\;
e^{-i(\omega-\omega^{\prime}) p\left[
t-z(t)\right] } \right. \nonumber
\\
&&+\rho^2\left. e^{-i(\omega-\omega^{\prime}) \left[
t-z(t)\right] }
\right. \nonumber
\\
&& \left.-\left[ (z,\dot{z},p,\rho) \rightarrow (-z,-\dot{z},r,\lambda)\right]\right\}\;. 
\nonumber 
\end{eqnarray}
We see that $F_{vac}$ and $F_{\left\langle \hat{a}^{\dag}\hat{a}\right\rangle}$ 
depend on $|\gamma|^2$, which has the same value for Dirichlet and Neumann
conditions. On the other hand $F_{\left\langle \hat{a}\hat{a}\right\rangle}$
depends on $\gamma^{*\;2}$ ( or $\gamma^{2}$), which differs by a sign in
Dirichlet and Neumann cases. Noting that the term 
${\cal C}_{\left\langle \hat{a}\hat{a}\right\rangle}$ (${\left\langle \hat{a}\hat{a}\right\rangle}\neq 0$)
is non-symmetric under time translations,
we can generalize the result found in Ref. \cite{Alves-Farina-Maia-Neto-JPA-2003},
concluding that for a general (relativistic) law of motion, 
Dirichlet and Neumann boundary conditions yield the same radiation force on a moving mirror when
the initial field state is symmetric under time translations.
In the non-relativistic limit, we recover 
$F _{vac}\left(t\right)\approx{|\gamma|^2\dddot{z}}/{6\pi }$,
found in Ref. \cite{Ford-Vilenkin-PRD-1982} for Dirichlet, 
and in Ref. \cite{Alves-Farina-Maia-Neto-JPA-2003} for Neumann condition. 
%

\begin{figure}
\epsfig{file=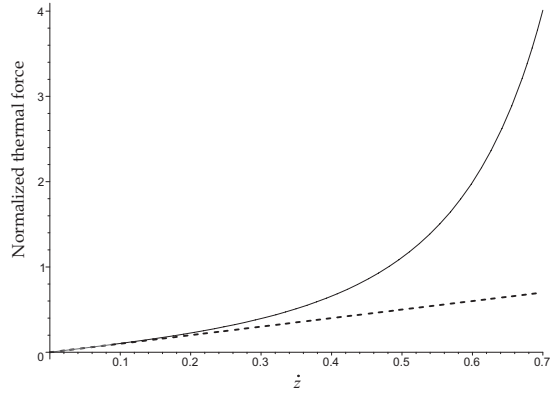,angle=00,width=0.4 \linewidth,clip=}
\caption{Normalized exact thermal force $|F^{(T)}|/\sigma_{T}$
(solid line) and approximate thermal force $|F_{(0)}^{(T)}|/\sigma_{T}$ (dashed line), both
valid for Dirichlet and Neumann boundary conditions, plotted as functions of
the mirror velocity $\dot z$.}
\label{force-temperature}
\end{figure}

Let us examine the radiation force on moving mirrors when
there are real particles in the initial state of the field.
We start with the thermal bath 
with temperature $T$, which is an example of 
invariant field state under time translations. 
For this state we need to take into account that
$
\left\langle \hat{a}_{\omega ^{\prime }}^{\dag }\hat{a}_{\omega
}\right\rangle =\bar{n}(\omega )\delta \left( \omega -\omega ^{\prime
}\right)  
$
where 
$
\bar{n}(\omega )={1}/({e^{\hbar \omega /T}-1}),
$
with the Boltzman constant equal to 1.
We get ${\cal T}_{\left\langle \hat{a}\hat{a}\right\rangle}=0$ and
${\cal T}_{\left\langle \hat{a}^{\dag}\hat{a}\right\rangle}$, renamed
as the energy density ${\cal T}_{\left\langle \hat{a}^{\dag}\hat{a}\right\rangle}^{(T)}$, 
is given by:
$
{\cal T}_{\left\langle \hat{a}^{\dag}\hat{a}\right\rangle}^{(T)}={|\gamma|^2\pi T^2}/
{12}\left[r^{\prime }\left( v\right) ^{2}
+p^{\prime }\left( u\right) ^{2}\right].
$
The force $F_{\left\langle \hat{a}\hat{a}\right\rangle}=0$, whereas
$F_{\left\langle \hat{a}^{\dag}\hat{a}\right\rangle}$, renamed
as the net thermal force $F^{(T)}$, 
is given by:
$$
F^{(T)}=-\sigma_{T}\left[\dot{
z}\dfrac{\left( 1+\dot{z}^{2}\right) }{\left( 1-\dot{z}^{2}\right) ^{2}}\right]
=-\sigma_{T}\sum_{n=0}^{\infty}(2n+1)\dot{z}^{2n+1},
$$
where $\sigma_{T}={2|\gamma|^2\pi T^{2}}/{3}$ is the viscosity coefficient.
This exact formula is a generalization of the one obtained in Ref. 
\cite{Jaekel-Reynaud-JPI-and-PLA-1993}. For non-relativistic velocities related, for instance,
to mechanical motions of the mirror ($\dot{z}\sim 10^{-8}$) or to the simulation of the mirror motion by changing
the reflectivity of a semiconductor by irradiation from laser ($\dot{z}\sim 10^{-3}$) \cite{ruoso-2005},
the series can be truncated in $n=0$, leading to the approximate formula: 
$F^{(T)}\approx F_{(0)}^{(T)}=-{2|\gamma|^2\pi T^{2}}\dot{z}/{3}$,
in agreement with Ref. \cite{Jaekel-Reynaud-JPI-and-PLA-1993} (for Dirichlet), and also with
Ref. \cite{Alves-Farina-Maia-Neto-JPA-2003} (for Neumann boundary condition). In the nonrelativistic context,
this approximate formula
is in good agreement with the exact value. For relativistically moving
mirrors (for instance, relativistic electron beam may be an embodiment of
a relativistic mirror \cite{granatstein-1976}), corrections to the approximate
formula can become necessary, as it can be seen in Fig. \ref{force-temperature}.
%

%

Let us now consider the coherent state, as an example of a non invariant state under time translations. 
The coherent state of amplitude $\alpha$ is defined as an eigenstate of the annihilation operator:
$
\hat{a}_{\omega }\left\vert \alpha \right\rangle =\alpha \delta \left(
\omega -\omega _{0}\right) \left\vert \alpha \right\rangle,
\label{aniquilacao-coerente}
$
where $\alpha = \left\vert \alpha \right\vert \exp(i\theta)$ and
$\omega _{0}$ is the frequency of the excited mode.
For this case, we have ${\cal T}_{\left\langle \hat{a}^{\dag}\hat{a}\right\rangle}$
and ${\cal T}_{\left\langle \hat{a}\hat{a}\right\rangle}$ 
relabeled as ${\cal T}_{\left\langle \hat{a}^{\dag}\hat{a}\right\rangle}^{(\alpha)}$
and ${\cal T}_{\left\langle \hat{a}\hat{a}\right\rangle}^{(\alpha)}$ respectively, and given by:
$$
{\cal T}_{\left\langle \hat{a}^{\dag}\hat{a}\right\rangle}^{(\alpha)}=
{|\gamma|^2}/{(2\pi) } \omega _{0}\left\{
\left\vert \alpha \right\vert ^{2}\left[ r^{\prime }\left( v\right) ^{2}
+p^{\prime }\left( u\right) ^{2}%
\right]\right\},
$$
\begin{eqnarray}
{\cal T}_{\left\langle \hat{a}\hat{a}\right\rangle}^{(\alpha)}&=&
-\frac{\omega _{0}}{4\pi } \left\{
\alpha ^{2}\left[ \gamma^2 \left[ r^{\prime }\left( v\right) \right]
^{2}e^{-2i\omega _{0}r\left( v\right) }\right.\right.\nonumber
\\
&&\left.\left.+\gamma^{*\;2}\left[ p^{\prime }\left(
u\right) \right] ^{2}e^{-2i\omega _{0}p\left( u\right) }\right]
+ \;\mbox{c.c.} \right\}.\nonumber
\end{eqnarray}
The exact forces ${F}_{\left\langle \hat{a}^{\dag}\hat{a}\right\rangle}$ and
${F}_{\left\langle \hat{a}\hat{a}\right\rangle}$, relabeled 
as the coherent forces ${F}_{\left\langle \hat{a}^{\dag}\hat{a}\right\rangle}^{(\alpha)}$
and ${F}_{\left\langle \hat{a}\hat{a}\right\rangle}^{(\alpha)}$ respectively, are given by:
$$
{F}_{\left\langle \hat{a}^{\dag}\hat{a}\right\rangle}^{(\alpha)}=
-\frac{4|\gamma|^2}{\pi }
 \omega _{0}
 \left\vert \alpha \right\vert
^{2}{\dot{z}}{\left( 1+\dot{z}%
^{2}\right) }/{\left( 1-\dot{z}^{2}\right) ^{2}},
$$
\begin{eqnarray}
F_{\left\langle \hat{a}\hat{a}\right\rangle
}^{\left( \alpha \right) }&=&-\frac{\omega _{0}}{4\pi }\left\vert \alpha
\right\vert ^{2}e^{-2i\left( \omega _{0}t-\theta \right) } \nonumber
\\
&&\times\left\{ \gamma ^{2}%
\left[ e^{2i\omega _{0}z(t)}\left( \frac{1-{\dot z}}{1+{\dot z}}%
\right) ^{2}-e^{-2i\omega _{0}z(t)}\right]\right.\nonumber
\\
&&\left. -\gamma ^{\ast 2}\left[ \left( 
\frac{1+{\dot z}}{1-{\dot z}}\right) ^{2}e^{-2i\omega
_{0}z(t)}-e^{2i\omega _{0}z(t)}\right] \right\} + {\mbox{c.c.}}.\nonumber
\end{eqnarray}
If we consider simultaneously nonrelativistic velocities and small displacements
(in the sense considered in Ref. \cite{Alves-Farina-Maia-Neto-JPA-2003}), 
according to what is required by Ford-Vilenkin approach \cite{Ford-Vilenkin-PRD-1982}, 
the force $F^{(\alpha)}={F}_{\left\langle \hat{a}^{\dag}\hat{a}\right\rangle}^{(\alpha)}+
{F}_{\left\langle \hat{a}\hat{a}\right\rangle}^{(\alpha)}$ can be approximated as:
\begin{eqnarray}
F^{\left( \alpha \right) }&\approx& -\frac{2\omega _{0}}{\pi }
\left\vert \alpha \right\vert ^{2}\left\{ 2|\gamma |^{2}\dot{z}\left(
t\right) -\left( \gamma ^{2}+\gamma ^{\ast 2}\right)\right.\nonumber
\\ \nonumber
&&
\times \left. \left[ \cos \left(
2\omega _{0}t-2\theta \right) \dot{z}\left( t\right) -\sin \left( 2\omega
_{0}t-2\theta \right) \omega _{0}z\left( t\right) \right] \right\}. 
\end{eqnarray}
\begin{figure}
\epsfig{file=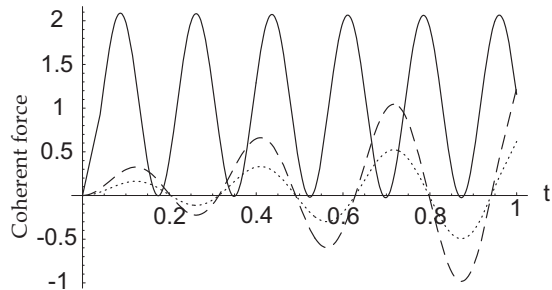,angle=00,width=0.4 \linewidth,clip=}
\caption{We plot the exact coherent force as function of time, for the Dirichlet
case. We consider $\alpha=\exp{(i\pi/2)}$, $\omega_0=10$,
and the following values of the velocity: $-10^{-8}$ (dotted line),
$-10^{-2}$ (dashed line) and $-8\times 10^{-1}$ (solid line).
The dotted and solid lines exhibit the coherent force
multiplied by the factors $5\times10^{5}$ and
1/125 respectively.}\label{force-coherent}
\end{figure}
In Fig. \ref{force-coherent} we plot the exact coherent force as a function of time 
for the Dirichlet boundary condition and
different values of the mirror velocity. 
Assuming the mirror moving with uniform velocity toward
the negative direction of the $x$-axis, the Fig. \ref{force-coherent}
shows the force oscillating $F^{(\alpha)}$ and the graph shifting to the
positive region of the vertical axis as the mirror velocity grows,
becoming the force more intense and opposite to the motion.
For the Neumann boundary condition, the force
oscillates in a different manner, but exhibits analogous shift
for relativistic velocities.

In summary, focusing on the advantages 
of the Fulling-Davies approach for the case of a massless scalar field in $1+1$ dimensions, 
in the present paper we showed the exact dynamical Casimir force
acting on a moving boundary under Neumann condition, with the vacuum as the initial state, 
generalizing the non-relativistic result found in Ref. \cite{Alves-Farina-Maia-Neto-JPA-2003}. For the thermal 
field, considering both Dirichlet and Neumann conditions, we wrote the
exact formula for the thermal force, generalizing the approximate
formula found in Ref. \cite{Jaekel-Reynaud-JPI-and-PLA-1993} and also, for instance,
in Refs. \cite{Jaekel-Reynaud-PLA-1993, plunien-PRL-2000,papers-temperatura-III,Alves-Farina-Maia-Neto-JPA-2003, Lambrecht-JOptB-2005}. 
For the coherent initial state, we found
exact formulas for the radiation force, which are different
if we consider Dirichlet or Neumann condition, generalizing
the perturbative formulas found in Ref. \cite{Alves-Farina-Maia-Neto-JPA-2003}.
Finally, we extended to a general (relativistic) law of motion
the conclusion found in the literature \cite{Alves-Farina-Maia-Neto-JPA-2003} that
Dirichlet and Neumann boundary conditions yield the same radiation force 
on a moving mirror when the initial field state is invariant under time translations.

We acknowledge L. C. B. Crispino and P. A. Maia Neto for careful reading 
of this paper. This work was supported by CNPq - Brazil.

%

\end{document}